\documentclass[12pt]{article}
\usepackage{graphicx}
\usepackage{latexsym}
\usepackage{amssymb}
\begin{document}
\hspace {-8mm} {\bf \large Accurate exchange energy and total energy for excited states: Inclusion of gradient correction.} \\ \\

\hspace {-8mm} {\bf Md.Shamim and Manoj K. Harbola}\\
{Department of Physics, Indian Institute of Technology,
Kanpur 208016, India}

\begin{abstract}
We present an approach for accurate calculation of exchange energy and total energy for excited states using time independent density functional formalism. This is done by inclusion of gradient correction into the excited state exchange energy functionals developed by us. We have incorporated Becke and Perdew Wang corrections into our functional and have studied various types of excited states having one and two gaps in occupation of orbitals.
\end{abstract}

\section{introduction}
Exchange energy functional for excited states which we reported in our earlier works \cite{sami2} does give accurate excitation energy but absolute values of total energy it leads to is not as accurate. Therefore, further correction needs to be made in this functional so as to obtain correct total energy of the sytem. Inclusion of gradient corrections to improve the accuracy of energy functionals has been in practice since the inception of density functional theory \cite{becke,PW,weiz}. In absence of exact energy functional for interacting electron systems, this is the only option one is left with. Such  corrections  are either derived from the properties of many electrons system or constructed impirically or semi-impirically. Generalized gradient approximation(GGA) and meta-generalized gradient approximation(meta-GGA) respectively are the examples of such corrections\cite{becke,PW}. While GGA uses gradient of density and its higher power, meta-GGA includes laplacian of density and its higer power. Perdew Wang \cite{PW} and Becke's \cite{becke} correction to exchange energy are the example of GGA and Perdew Burke Ernzerhof's \cite{pbe} correction to exchange energy is a meta-GGA kind of correction.

In Kohn-Sham formulation of density functional theory \cite{ks} kinetic energy part is well taken care of. And correlation  energy which does play important role in many instances is only a meager part. Therefore, it is basically the ecxhange energy which needs to be treated accurately at lease so far as we are concerned with calculation of total energy of the system. For ground states, the local density approximation for exchange energy functional gives an everage error of about 10 $\%$.  To minimize this error gradient correction for exchange has to be used. Among others, the widely used gradient corrections for exchange energy are due to Becke \cite{becke} and Perdew Wang \cite{PW}. Becke's correction is a semiempirical one and the PW's correction is obtained by gradient expansion of exchange hole.

\section{Gradient correction for excited state exchange energy functional}

Exchange energy functional reported in reference \cite{sami2} is quite accurate as we have seen in the results for excitation energies obtained by using it. To make it more accurate we are going to incorporate gradient corrections into it. Since doing gradient correction for excited is not as simple as for the ground state cases. Therefore we resort to some sort of approximation. We see that  Becke's and PW's corrections to exchange energy have been constructed to satisfy certain properites which are not specfic to the ground states. Such as the asymptotic decay of exchange energy density should go as $\frac{-1}{2r}$.  Therefore, we hope that these gradient correction will work equally well for excited states also. With this intuition we use gradient corrections for exchange energy due to Becke \ref{becke} and Perdew Wang(PW) \ref{PW} to calculate exchange energy and total energy for excited states..

Exchange energy functional for excited states with Becke's gradient correction is given as,
\begin{equation}
E_{X}^{MLSD(Becke)} = E_{X}^{MLSD} - \beta \sum_{\sigma}\int \rho_{\sigma}^{4/3} \frac{x_{\sigma}^2}{1+6\beta x_{\sigma}sinh^{-1}x_{\sigma}}d{\bf r}
\label{beckeex}
\end{equation}
where, $x_{\sigma} = \frac{\nabla \rho_{\sigma}}{\rho_{\sigma}^{4/3}}$ and $\beta = 0.0042$
 and  exchange energy functional for excited states with PW's gradient correction is given as
\begin{equation}
E_{X}^{MLSD(PW)} = \int\epsilon_{X}^{MLSD} F(s)d{\bf r}
\label{pwex}
\end{equation}
where,
\begin{eqnarray}
\epsilon_{X}^{MLSD} &=& \rho({\bf r})\left[\epsilon(k_{3}) - \epsilon(k_{2})+ \epsilon(k_{1})\right]
+ \frac{1}{8\pi^{3}}(k_{3}^{2}-k_{1}^{2})^{2}\;ln\left(\frac{k_{3}+k_{1}}{k_{3}-k_{1}}
\right) \nonumber \\
&-&\frac{1}{8\pi^{3}}(k_{3}^{2}-k_{2}^{2})^{2}\;ln\left(\frac{k_{3}+k_{2}}{k_{3}-k_{2}}\right) 
-\frac{1}{8\pi^{3}}(k_{2}^{2}-k_{1}^{2})^{2}\;ln\left(\frac{k_{2}+k_{1}}{k_{2}-k_{1}}\right)
\label{eq:eps}
\end{eqnarray}
and, $s= \frac{\nabla \rho}{2k_f \rho}$, $F= \left( 1+ 0.086s^2/m + bs^4 + cs^6\right)^m$ with $m=1/15, b= 14, c=0.2 $

\section{Results}

We have calculated total energies and excitation energies of excited states of various atoms using these gradient corrections for  exchange enrgies. The results obtained in this way are very encouraging, giving total energy for excited states which are very close to the corresponging exact exchange only values. In the tables \ref{TE1} to \ref{TE6}  we present total energies and excitation of excited states obtained with Becke and PW corrections along with the exact exchange only (Hartree-Fock(HF)) values.

To calculate exchange energy and total energy we do Kohn Sham calculation with LSD approximation for exchange potential. From the orbitals generated in this way we calculate the $k$ wave vectors from which we get exchage energy. We use this exchange energy to get total energy for the atoms or ions.  

\begin{table}
\caption{\label{TE1}Total energy and excitation energy of various atoms/ions. }
\vspace{0.2in}
\begin{tabular}{lccccccc}
\hline
atoms/ions & -$E_{HF}^{\ast}$ &- $E_{Becke}^{\ast}$ & -$E_{PW}^{\ast}$ & $\Delta E_{HF}$ &
$\Delta E_{Becke}$ & $ \Delta E_{PW}$  \\
\hline
$N(2s^{2}2p^{3}\;^{4}S\rightarrow2s^{1}2p^{4}\;^{4}P)$ &53.988&53.989&54.026&0.413&0.409&0.423 \\
$O^{+}(2s^{2}2p^{3}\;^{4}S\rightarrow2s^{1}2p^{4}\;^{4}P)$ &73.820&73.797&73.835&0.553&0.564&0.582 \\
$O(2s^{2}2p^{4}\;^{3}P\rightarrow2s^{1}2p^{5}\;^{3}P)$ &74.184&74.181&74.227&0.625&0.631&0.652 \\
$F^{+}(2s^{2}2p^{4}\;^{3}P\rightarrow2s^{1}2p^{5}\;^{3}P)$ &98.033&98.006&98.051&0.799&0.802&0.0.840 \\
$F(2s^{2}2p^{5}\;^{2}P\rightarrow2s^{1}2p^{6}\;^{2}S)$ &98.531&98.561&98.610&0.878&0.869&0.898 \\
$Ne^{+}(2s^{2}2p^{5}\;^{2}P\rightarrow2s^{1}2p^{6}\;^{2}S)$ &126.861&126.745&126.791&1.083&1.074&1.111 \\
\hline
\end{tabular}
\end{table}

\begin{table}
\caption{\label{TE2}Total energy and excitation enery of various atoms/ions.}
\vspace{0.2in}
\begin{tabular}{lcccccc}
\hline
atoms/ions & -$E_{HF}^{\ast}$ &- $E_{Becke}^{\ast}$ & -$E_{PW}^{\ast}$ & $\Delta E_{HF}$ &
$\Delta E_{Becke}$ & $ \Delta E_{PW}$  \\
\hline
$Li(2s^{1}\;^{2}S\rightarrow2p^{1}\;^{2}P)$& 7.365& 7.357&7.366 &0.068&0.0702&0.075 \\
$Na(3s^{1}\;^{2}S\rightarrow3p^{1}\;^{2}P)$ &161.786&161.804&161.890&0.073&0.0763&0.083\\
$Mg^{+}(3s^{1}\;^{2}S\rightarrow3p^{1}\;^{2}P)$ &199.214&199.210&199.297&0.158&0.171&0.182 \\
\hline
\end{tabular}
\end{table}

\begin{table}
\caption{\label{TE3}Total energy and excitation enery of various atoms/ions.}
\vspace{0.2in}
\begin{tabular}{lcccccc}
\hline
atoms/ions & -$E_{HF}^{\ast}$ &- $E_{Becke}^{\ast}$ & -$E_{PW}^{\ast}$ & $\Delta E_{HF}$ &
$\Delta E_{Becke}$ & $ \Delta E_{PW}$  \\
\hline
$P(3s^{2}3p^{3}\;^{4}S\rightarrow3s^{1}3p^{4}\;^{4}P)$ &340.417&340.400&340.507&0.302&0.307&0.324 \\
$S(3s^{2}3p^{4}\;^{3}P\rightarrow3s^{1}3p^{5}\;^{3}P)$ &397.079&397.053&397.165&0.426&0.436&0.454 \\
$Cl^{+}(3s^{2}3p^{4}\;^{3}P\rightarrow3s^{1}3p^{5}\;^{3}P)$ &458.523&458.484&458.596&0.526&0.544&0.564 \\
$Cl(3s^{2}3p^{5}\;^{2}P\rightarrow3s^{1}3p^{6}\;^{2}S)$ & 458.917&458.900&459.015&0.565&0.567&0.585\\
$Ar^{+}(3s^{2}3p^{5}\;^{2}P\rightarrow3s^{1}3p^{6}\;^{2}S)$ &525.598&525.574&525.686&0.677&0.671&0.703 \\
\hline
\end{tabular}
\end{table}

\begin{table}
\caption{\label{TE4}Total energy and excitation enery of various atoms/ions.}
\vspace{0.2in}
\begin{tabular}{lcccccc}
\hline
atoms/ions & -$E_{HF}^{\ast}$ &- $E_{Becke}^{\ast}$ & -$E_{PW}^{\ast}$ & $\Delta E_{HF}$ &
$\Delta E_{Becke}$ & $ \Delta E_{PW}$  \\
\hline
$P(2s^{2}3p^{3}\;^{4}S\rightarrow2s^{1}3p^{4}\;^{4}P)$ &333.834&333.708&333.766&6.882&7.000&7.065 \\
$S(2s^{2}3p^{4}\;^{3}P\rightarrow2s^{1}3p^{5}\;^{3}P)$ &389.257&389.116&389.174&8.246&8.373&8.445 \\
$Cl^{+}(2s^{2}3p^{4}\;^{3}P\rightarrow2s^{1}3p^{5}\;^{3}P)$ &449.234&449.079&449.132&9.812&9.949&10.028 \\
$Cl(2s^{2}3p^{5}\;^{2}P\rightarrow2s^{1}3p^{6}\;^{2}S)$ &449.768&449.602&449.659&9.714&9.864&9.941\\
$Ar^{+}(2s^{2}3p^{5}\;^{2}P\rightarrow2s^{1}3p^{6}\;^{2}S)$ &514.882&514.699&514.751&11.393&11.566&11.638 \\
\hline
\end{tabular}
\end{table}

\begin{table}
\caption{\label{TE5}Total energy and excitation enery of various atoms/ions.}
\vspace{0.2in}
\begin{tabular}{lcccccc}
\hline
atoms/ions & -$E_{HF}^{\ast}$ &- $E_{Becke}^{\ast}$ & -$E_{PW}^{\ast}$ & $\Delta E_{HF}$ &
$\Delta E_{Becke}$ & $ \Delta E_{PW}$  \\
\hline
$B(2s^{2}2p^{1}\;^{2}P\rightarrow2s^{1}2p^{2}\;^{2}D)$ &24.312&24.304&24.330&0.217&0.211&0.220 \\
$C^{+}(2s^{2}2p^{1}\;^{2}P\rightarrow2s^{1}2p^{2}\;^{2}D)$ &36.963&36.945&36.975&0.329&0.328&0.342 \\
$C(2s^{2}2p^{2}\;^{3}P\rightarrow2s^{1}2p^{3}\;^{3}D)$ &37.395&37.376&37.409&0.294&0.303&0.315 \\
$N^{+}(2s^{2}2p^{2}\;^{3}P\rightarrow2s^{1}2p^{3}\;^{3}D)$ & 53.917&53.434&53.471&0.414&0.439&0.454\\
$Si^{+}(3s^{2}3p^{1}\;^{2}P\rightarrow3s^{1}3p^{2}\;^{2}D)$ & 288.299&288.289&288.394&0.274&0.282&0.297\\
$Si(3s^{2}3p^{2}\;^{3}P\rightarrow3s^{1}3p^{3}\;^{3}D)$ & 288.62&288.606&288.711&0.234&0.246&0.260\\
\hline
\end{tabular}
\end{table}

\begin{table}
\caption{\label{TE6}Total energy and excitation enery of various atoms/ions.}
\vspace{0.2in}
\begin{tabular}{lcccccc}
\hline
atoms/ions & -$E_{HF}^{\ast}$ &- $E_{Becke}^{\ast}$ & -$E_{PW}^{\ast}$ & $\Delta E_{HF}$ &
$\Delta E_{Becke}$ & $ \Delta E_{PW}$  \\
\hline
$Be(2s^{2}\;^{1}S\rightarrow2p^{2}\;^{1}D)$& 14.301&14.291&14.300&0.272&0.264&0.288\\
$B(2s^{2}2p^{1}\;^{2}P\rightarrow2p^{3}\;^{2}D)$ &24.059&24.023&24.037&0.470&0.491&0.513\\
$C^{+}(2s^{2}2p^{1}\;^{2}P\rightarrow2p^{3}\;^{2}D)$ & 36.595&36.540&36.552&0.697&0.733&0.765\\
$C(2s^{2}2p^{2}\;^{3}P\rightarrow2p^{4}\;^{3}P)$ & 36.946&36.934&36.948&0.743&0.746&0.776\\
$N^{+}(2s^{2}2p^{2}\;^{3}P\rightarrow2p^{4}\;^{3}P)$ & 52.865&52.841&52.892&1.023&1.032&1.173\\
$N(2s^{2}2p^{3}\;^{4}S\rightarrow2p^{5}\;^{2}P)$ &53.222&53.202&53.224&1.179&1.196&1.225\\
$O^{+}(2s^{2}2p^{3}\;^{4}S\rightarrow2p^{5}\;^{2}P)$ & 72.829&72.795&72.811&1.544&1.566&1.606\\
$O(2s^{2}2p^{4}\;^{3}P\rightarrow2p^{6}\;^{1}S)$ &73.306&73.319&73.339&1.5032&1.494&1.540\\
$F^{+}(2s^{2}2p^{4}\;^{3}P\rightarrow2p^{6}\;^{1}S)$ & 96.934&96.935&96.947&1.898&1.883&1.944\\
$Mg(3s^{2}\;^{1}S\rightarrow3p^{2}\;^{1}D)$ & 199.357&199.362&199.480&0.258&0.257&0.255\\
$S(3s^{2}3p^{4}\;^{3}P\rightarrow3p^{6}\;^{1}S)$ & 396.478&96.455&396.552&1.027&1.034&1.067\\
$P(3s^{2}3p^{3}\;^{4}S\rightarrow3p^{5}\;^{2}P)$ &339.865&339.834&339.930&0.854&0.874&0.901\\
$Si^{+}(3s^{2}3p^{1}\;^{2}P\rightarrow3p^{3}\;^{2}D)$ & 287.987&287.944&288.034&0.586&0.627&0.657\\
$Si(3s^{2}3p^{2}\;^{3}P\rightarrow3p^{4}\;^{3}P)$ & 288.268&288.250&288.342&0.586&0.602&0.629\\
$Cl^{+}(3s^{2}3p^{2}\;^{3}P\rightarrow3p^{4}\;^{3}P)$ & 457.795&457.767&457.860&1.254&1.261&1.300\\
\hline
\end{tabular}
\end{table}

\section{Two gap systems}

We have applied the gradient corrected exchange energy functional to excited states with two-gaps also. In these cases we use the exchage energy functional for two-gap systems has the same form as the one-gap case but has more terms. For details the readers may refer \cite{sami2}.  The results calculated for total energy and excitation energies with Becke and PW corrections are given in tables \ref{TE2g} along with HF values. 

\begin{table}
\caption{\label{TE2g}Total energy and excitation enery of various atoms/ions having two gaps.}
\vspace{0.2in}
\begin{tabular}{lcccccc}
\hline
atoms/ions & -$E_{HF}^{\ast}$ &- $E_{Becke}^{\ast}$ & -$E_{PW}^{\ast}$ & $\Delta E_{HF}$ &
$\Delta E_{Becke}$ & $ \Delta E_{PW}$  \\
\hline
$B(2p^{1}3p^{2}\;M_L=3,M_S=1/2)$ & 23.516&23.441&23.442&1.010&1.064&1.108\\
$N(2p^{4}3p^{1}\;M_L=2,M_S=3/2)$ & 52.958&52.874&52.884&1.438&1.524&1.565\\
$F(2p^{6}3p^{1}\;M_L=1,M_S=1/2)$ &97.034& 96.959&96.971&2.371&2.471&2.527\\
$Ne(2p^{6}3p^{2}\;M_L=1,M_S=1)$ & 124.594&124.420&124.424&3.948&4.167&4.248\\
$Na(2s^{1}2p^{6}3p^{2}\;M_L=1,M_S=1/2)$ & 159.236&159.159&159.203&2.615&2.722&2.770\\
$Al(2s^{1}2p^{6}3s^{1}3p^{3}\;M_L=0,M_S=1/2)$& 237.16&237.064&237.115&4.702&4.816&4.879\\ 
$Ar(2p^{6}3s^{0}3p^{6}4p^{2}\;M_L=2,M_S=0)$& 524.215&524.074&524.156&2.589&2.722&2.776\\ 
$K(2p^{6}3s^{1}3p^{6}4p^{2}\;M_L=1,M_S=1/2)$& 597.523&597.436&597.542&1.626&1.850&2.169\\ 
\hline
\end{tabular}
\end{table}

\section{Concluding remarks and future prospects}
We have added gradient correction to exchange energy functional to obtain better approximation for exchange energy functional.
 The correction term used here have been borrowed from ground state DFT on some physical grounds and are found to lead the
 result in right direction as expected. 
 Becke and PW corrections which we have employed here are found to consistently improve the results for excited states exchange
 energy, total energy and excitation energy. The results obtained from such calculation compare very closely with Hartree Fock
 results.

\newpage

\end{document}